\documentclass[twocolumn,aps,pr,superscriptaddress,preprintnumbers,nofootinbib,10pt]{revtex4-2}
\usepackage{multirow}
\usepackage{amsmath}
\usepackage{amssymb}
\usepackage[dvipdf,dvips]{graphicx}
\usepackage{color}
\usepackage{hyperref}
\usepackage{url}
\usepackage{slashed}
\usepackage{subfigure}
\usepackage[usenames,dvipsnames]{xcolor}
\usepackage{amsmath}
\usepackage{amsfonts}
\usepackage{float} 
\usepackage{amssymb}
\usepackage{epsfig}
\usepackage{graphics}
\usepackage{euscript}
\usepackage{slashed}
\usepackage{epstopdf}
\usepackage[utf8]{inputenc}
\allowdisplaybreaks
\usepackage[normalem]{ulem}
\usepackage{pifont}
\usepackage{dsfont}
\usepackage{MnSymbol}
\usepackage{verbatim}
\usepackage{graphicx}
\usepackage{latexsym}
\usepackage{tikz-feynman}

\begin{document}
	
	\title{Weyl operators, Tomita-Takesaki theory, and Bell-CHSH inequality violations}
	
	\author{P. De Fabritiis} \email{pdf321@cbpf.br} \affiliation{CBPF $-$ Centro Brasileiro de Pesquisas Físicas, Rua Dr. Xavier Sigaud 150, 22290-180, Rio de Janeiro, Brazil}
	
	\author{F. M. Guedes} \email{fmqguedes@gmail.com} \affiliation{UERJ $–$ Universidade do Estado do Rio de Janeiro,	Instituto de Física $–$ Departamento de Física Teórica $–$ Rua São Francisco Xavier 524, 20550-013, Maracanã, Rio de Janeiro, Brazil}
	
	\author{M. S.  Guimaraes}\email{msguimaraes@uerj.br} \affiliation{UERJ $–$ Universidade do Estado do Rio de Janeiro,	Instituto de Física $–$ Departamento de Física Teórica $–$ Rua São Francisco Xavier 524, 20550-013, Maracanã, Rio de Janeiro, Brazil}
	
	\author{G. Peruzzo} \email{gperuzzofisica@gmail.com} \affiliation{UFF $-$ Instituto de F\'{i}sica, Universidade Federal Fluminense, Campus da Praia Vermelha, Av. Litor\^{a}nea s/n, 24210-346, Niter\'{o}i, RJ, Brazil}
	
	\author{I. Roditi} \email{roditi@cbpf.br} \affiliation{CBPF $-$ Centro Brasileiro de Pesquisas Físicas, Rua Dr. Xavier Sigaud 150, 22290-180, Rio de Janeiro, Brazil}
	
	\author{S. P. Sorella} \email{silvio.sorella@gmail.com} \affiliation{UERJ $–$ Universidade do Estado do Rio de Janeiro,	Instituto de Física $–$ Departamento de Física Teórica $–$ Rua São Francisco Xavier 524, 20550-013, Maracanã, Rio de Janeiro, Brazil}

	\begin{abstract}
The violation of the Bell-CHSH inequality in the vacuum state of a relativistic free real scalar field is established by means of the Tomita-Takesaki construction and of the direct computation of the correlation functions of Weyl operators. 
	\end{abstract}

	\maketitle	
	
	\section{Introduction}

Nowadays, the study of entanglement properties within the relativistic Quantum Field Theory framework is a  challenging and fast growing area due, in particular, to the many potential applications to quantum information theory~\cite{Witten:2018zxz, Faulkner:2022mlp}. As underlined in~\cite{Witten:2018zxz},   to investigate entanglement in Quantum Field Theory, the handling of the algebra of observables is as important as to deal with states.  This hints towards the relevance of Tomita-Takesaki's theory to properly address those concepts. 

Among the various quantities introduced to study entanglement,  the Bell-Clauser-Horne-Shimony-Holt inequality~\cite{Bell:1964kc,Clauser:1969ny} plays a prominent role. In particular, the search for Bell-CHSH inequality violations in elementary particle processes at high-energies is receiving great attention both from phenomenological and experimental points of view, as one can see in~\cite{Fabbrichesi21, Severi22, Afik21, Afik22, Afik23, Barr22,Ashby-Pickering:2022umy} and references therein. From a more theoretical side, to investigate Bell-CHSH inequalities requires to face fundamental aspects of Quantum Field Theory such as causality, the vacuum state nature, and the role of localized operator algebras in Minkowski spacetime. Here we quote the seminal works~\cite{Summers:1987fn,Summ,Summers:1987ze} where the authors, by using Algebraic Quantum Field Theory techniques~\cite{Haag:1992hx}, have been able to prove the existence of Bell-CHSH inequality violations in the vacuum even at the level of free fields. Other studies of the Bell-CHSH inequality in Quantum Field Theory can be found in~\cite{Summers:1995mv,Verch}.

More recently, we investigated further aspects of the Quantum Field Theory formulation of Bell-CHSH inequalities~\cite{Peruzzo:2022pwv,Peruzzo:2022tog,Sorella:2023pzc,DeFabritiis:2023ubj,Dudal:2023pbc,Dudal:2023mij}. For instance, the Bell-CHSH inequality was studied in~\cite{Peruzzo:2022pwv} for a pair of free massive scalar fields; in~\cite{Peruzzo:2022tog}, one can find the Feynman path integral formulation of this inequality; the relationship between Bogoliubov transformations, squeezed states and entanglement was discussed in~\cite{Sorella:2023pzc}. The generalization of Mermin's inequalities to spinor Quantum Field Theory has been investigated in~\cite{DeFabritiis:2023ubj}; the BRST invariance of the Bell-CHSH inequality in gauge theories using the Kugo-Ojima analysis was discussed in~\cite{Dudal:2023pbc}; the explicit construction of test functions leading to the Bell-CHSH inequality violation using Haar wavelets and the Planck-taper window function has been achieved in~\cite{Dudal:2023mij} in the context of the $1+1$ Thirring model. 

The present work aims to give continuity to our previous efforts. More precisely, we shall provide a study of the Bell-CHSH inequality in the vacuum state of a real free scalar massive field in $1+3$ Minkowski spacetime by making  use of the Weyl operators,  which play an important role in order to construct Von Neumann algebras of field operators localized in suitable spacetime regions.

This work is organized as follows. In Section~\eqref{Sect2}, we give a brief reminder of canonical quantization for a real scalar field in order to fix the notation. In Section~\ref{Sect3}, we provide an account of the basic properties of Weyl operators and algebras, introducing  the main  concepts that will be used throughout this work. The Bell-CHSH inequality in the Quantum Field Theory framework is introduced in Section~\ref{Sect4}, adopting an Hermitean combination of Weyl operators to perform the role of Alice's and Bob's operators, being the final result for the Bell-CHSH correlation expressed in terms of the scalar products between the test functions needed to properly localize the field operators in spacelike separated regions. In Section~\ref{Sect5} we discuss the Tomita-Takesaki theory,  outlining the construction of Bob's operators from Alice's ones by means of the  conjugation operator. The careful choice of test functions leading to the explicit Bell-CHSH inequality violation is done in Section~\ref{Sect6}. To some extent, this choice of test functions can be compared to choosing the free angle entering the Bell-CHSH inequality of Quantum Mechanics. We state our conclusions in Section~\ref{Sect7}. For reader's benefit, Appendix~\ref{A} contains a concise summary of basic notions on Von Neumann algebras and on the Tomita-Takesaki construction~\cite{BR}, while Appendix~\ref{B} is devoted to a short summary on a few basic elements of the spectral properties of self-adjoint operators.  
\newpage

\section{Quantum Field Theory basics}\label{Sect2}
	
Let us start by recalling  a  few  properties of the canonical quantization of a free massive scalar field \cite{Haag:1992hx}:  
\begin{equation} 
{\cal L} =   \frac{1}{2} \left( \partial^\mu \varphi \partial_\mu \varphi - m^2 \varphi^2 \right).  \label{cnq1}
\end{equation} 
Expanding the field $\varphi$ in terms of creation and annihiliation operators, one gets 
\begin{equation} \label{qf}
\varphi(t,{\vec x}) = \int \frac{d^3 {\vec k}}{(2 \pi)^3} \frac{1}{2 \omega_k} \left( e^{-ikx} a_k + e^{ikx} a^{\dagger}_k \right), 
\end{equation} 
with $\omega_k  = k^0 = \sqrt{{\vec{k}}^2 + m^2}$ and
\begin{align}\label{ccr}
[a_k, a^{\dagger}_q] &= (2\pi)^3 2\omega_k \delta^3({\vec{k} - \vec{q}}), \\ \nonumber 
[a_k, a_q] &= [a^{\dagger}_k, a^{\dagger}_q] = 0, 
\end{align}
are the canonical commutation relations. From the above definition, one can show that
\begin{equation} 
\left[ \varphi(x) , \varphi(y) \right] = i \Delta_{\textrm{PJ}} (x-y), \label{caus} 
\end{equation}
where $\Delta_{\textrm{PJ}}(x-y) $ is the Lorentz-invariant causal Pauli-Jordan function, defined by (here, $\varepsilon(x) \equiv \theta(x) - \theta(-x)$):
\begin{align}\label{it1}
	 i \Delta_{\textrm{PJ}}(x-y) \!=\!\! \int \!\! \frac{d^4k}{(2\pi)^3} \varepsilon(k^0) \delta(k^2-m^2) e^{-ik(x-y)}. 
\end{align}
It is well-known that the Pauli-Jordan function vanishes when $(x-y)^2 < 0$, thus encoding the principle of relativistic causality. Furthermore, the Pauli-Jordan function satisfies $(\partial^2_x + m^2) \Delta_{\textrm{PJ}}(x-y) = 0$ and $\Delta_{\textrm{PJ}}(x-y) = - \Delta_\textrm{{PJ}}(y-x)$. Explicitly, it can be rewritten as
 \begin{align} 
 	\Delta_{\textrm{PJ}}(x-y) &= \left[ \frac{\theta(x^0-y^0) -\theta(y^0-x^0)}{2\pi} \right] \times \left[  -\delta\left((x-y)^2\right)   \right. \nonumber \\
 	& \left.  +m \frac{\theta((x-y)^2)  J_1(m\sqrt{(x-y)^2}) }{ 2 \sqrt{(x-y)^2}}\right],
 \end{align} 
 where $J_1$ is the Bessel function.

The quantum fields are too singular objects, being in fact operator-valued distributions~\cite{Haag:1992hx}. As such, they have to be smeared out in order to provide well defined operators acting in the Hilbert space, namely,
\begin{equation} \label{sm}
\varphi(h) = \int d^4x \;\varphi(x) h(x).
\end{equation} 
In the above expression, $h(x)$ is a test function belonging to the space of compactly supported smooth functions ${\cal C}_{0}^{\infty}(\mathbb{R}^4)$.  Defining the Fourier transform as   
\begin{equation}\label{fft}
{\hat h}(p) = \int d^4x \; e^{ipx} h(x), 
\end{equation} 
and plugging expression~\eqref{qf} in~\eqref{sm}, we find 
\begin{equation} \label{smft}
\varphi(h) = \int \frac{d^3 {\vec k}}{(2 \pi)^3} \frac{1}{2 \omega_k} \left( {\hat h}^{*}(\omega_k,{\vec k}) a_k + {\hat h}(\omega_k,{\vec k}) a^{\dagger}_k \right).  
\end{equation}
This can be immediately rewritten as
\begin{align}\label{key}
	\varphi(h) = a_h + a^{\dagger}_h,
\end{align}
if we define the smeared versions for the creation and annihilation operators  as
\begin{align} 
a_h &= \int \frac{d^3 {\vec k}}{(2 \pi)^3} \frac{1}{2 \omega_k}  {\hat h}^{*}(\omega_k,{\vec k}) a_k, \nonumber \\
a^{\dagger}_h &= \int \frac{d^3 {\vec k}}{(2 \pi)^3} \frac{1}{2 \omega_k} {\hat h}(\omega_k,{\vec k}) a^{\dagger}_k. 
\end{align} 
The canonical commutation relations now read
\begin{equation} 
\left[ a_h, a^{\dagger}_{h'}\right]  = \langle h \vert h' \rangle, \label{ccrfg}
\end{equation}
with $ \langle h \vert h' \rangle$ denoting the Lorentz invariant scalar product between the test functions $h$ and $h'$, {\it i.e.},
\begin{align} 
\langle h \vert h' \rangle &= \int \frac{d^3 {\vec k}}{(2 \pi)^3} \frac{1}{2 \omega_k}  {\hat h}^{*}(\omega_k,{\vec k}) {\hat h'}(\omega_k, {\vec k}) \nonumber \\
&= 
\int \frac{d^4 {\vec k}}{(2 \pi)^4} 2\pi \;\theta(k^0) \delta(k^2-m^2)  {\hat h}^{*}(k) {\hat h}'(k). \label{scpd}
\end{align} 
This scalar product can be recast in configuration space
\begin{equation} 
\langle h \vert h' \rangle = \int d^4x d^4x'\; h(x) {\cal D}(x-x') h'(x'), \label{confsp} 
\end{equation} 
where ${\cal D}(x-x')$ is the so-called Wightman function
\begin{equation} 
{\cal D}(x-x') = \langle 0 \vert \varphi(x) \varphi(x')  \vert 0 \rangle \!=\!\!\! \int \!\!\! \frac{d^3 {\vec k}}{(2 \pi)^3} \frac{1}{2 \omega_k} e^{-ik(x-x')}, \label{Wg}
\end{equation} 
which can be decomposed as 
\begin{equation} 
{\cal D}(x-x') = \frac{i}{2} \Delta_{\textrm{PJ}}(x-x')   + H(x-x'), \label{decomp}
\end{equation} 
being $H(x-x') = H(x'-x)$ the real symmetric quantity
\begin{equation} 
H(x-x') = \frac{1}{2} \int \frac{d^3 {\vec k}}{(2 \pi)^3} \frac{1}{2 \omega_k} \left( e^{-ik(x-x')} + e^{ik(x-x')}
 \right). \label{H}
\end{equation} 
The commutation relation~\eqref{caus} can be expressed in terms of the smeared quantum fields as
\begin{equation} 
\left[ \varphi(h) , \varphi(h') \right] =i  \Delta_{\textrm{PJ}}(h,h'), \label{comm_smeared}
\end{equation}
where $(h , h')$ are test functions and
\begin{equation} 
\Delta_{\textrm{PJ}}(h,h')= \int d^4x \; d^4x' h(x) \Delta_{\textrm{PJ}}(x-x')  h'(x'). \label{pauli_jordan_smeared}
\end{equation}
Therefore, the causality condition in terms of smeared fields can be recast as: if the supports of $h$ and $h'$ are spacelike separated, then we have $\left[ \varphi(h) , \varphi(h') \right] = 0$.

\section{Weyl operators and algebras}\label{Sect3}

In this section, we introduce the main algebraic concepts that will be used throughout this work. 
Let ${\cal O}$ stand for an open region of the Minkowski spacetime and let ${\cal M}({\cal O})$ be the space of test functions in ${\cal C}_{0}^{\infty}(\mathbb{R}^4)$ with support contained in $\cal O$: 
\begin{equation} 
	{\cal M}({\cal O}) = \{ f; \; {\rm such \; that} \; supp(f) \subseteq {\cal O} \}. \label{MO}
\end{equation}
Following \cite{Summers:1987fn,Summ}, we introduce the symplectic complement of ${\cal M}({\cal O})$ as 
\begin{equation} 
	{\cal M'}({\cal O}) = \{ g; \; \Delta_{\textrm PJ}(g,f) =0, \;\; \forall f \in {\cal M}({\cal O})\}. \label{MpO}
\end{equation}
That is, ${\cal M'}({\cal O})$ is given by the set of all test functions for which the smeared Pauli-Jordan  expression defined by Eq.~\eqref{pauli_jordan_smeared} vanishes for any $f$ belonging to ${\cal M}({\cal O})$. The usefulness of the symplectic complement ${\cal M'}({\cal O})$ relies on the fact that it allows us to rephrase causality as 
\begin{equation} 
	\left[ \varphi(g) , \varphi(f) \right] = 0 \;, \;\; \forall g\in {\cal M'}({\cal O}) \;\;{\rm and} \;\; \forall f \in {\cal M}({\cal O}). \label{MMM}
\end{equation}

Let us introduce the  Weyl operators,  an important class of unitary operators obtained by exponentiating the smeared field, namely 
\begin{equation} 
{\cal A}_h = e^{i {\varphi}(h) }\;, \label{Weyl}
\end{equation}
with ${\varphi}(h)$ defined in Eq.~\eqref{sm}.  Using the relation
\begin{equation}
e^A \; e^B = \; e^{ A+B +\frac{1}{2}[A,B]}, \label{exp_AB}
\end{equation} 
valid for two operators $(A,B)$ commuting with $[A,B]$, one  checks that the Weyl operators give rise to the following algebraic structure:
\begin{align} \label{algebra} 
	{\cal A}_f {\cal A}_g   &= e^{ - \frac{i}{2} \Delta_{\textrm{PJ}}(f, g)}\;{\cal A}_{(f+g)}, \nonumber \\
	{\cal A}_f^\dagger {\cal A}_f &= 	{\cal A}_f {\cal A}_f^\dagger = 1, \nonumber \\ 
	{\cal A}^{\dagger}_f &=  {\cal A}_{(-f)}	 
\end{align} 
where $\Delta_{\textrm{PJ}}(f,g)$ is the smeared Pauli-Jordan expression~\eqref{pauli_jordan_smeared}. In particular, we can  see that when $\Delta_{\textrm{PJ}}(f,g)$ vanishes, we find ${\cal A}_f {\cal A}_g = {\cal A}_g {\cal A}_f$, that is, the Weyl operators commute. Expressing the smeared quantum field as $\varphi(h) = a_h + a_h^\dagger$ and using equations~\eqref{ccrfg} and \eqref{exp_AB}, one can compute the vacuum expectation value of the Weyl operator ${\cal A}_h$, finding 
\begin{equation} 
\langle 0 \vert  {\cal A}_h  \vert 0 \rangle = \; e^{-\frac{1}{2} {\lVert h\rVert}^2}, \label{vA}
\end{equation} 
with ${\lVert h\rVert}^2 \equiv \langle h | h \rangle$ and the vacuum state $\vert 0 \rangle$ is the Fock vacuum defined by the condition $a_k \vert 0 \rangle=0, \forall k$. In particular, if $supp_f$ and $supp_g$ are spacelike separated, causality ensures that the Pauli-Jordan function vanishes. Thus, from the above properties, it follows the important relation 
\begin{equation} 
\langle 0 \vert  {\cal A}_f {\cal A}_{g}  \vert 0 \rangle =  \langle 0 \vert {\cal A}_{(f + g)} \vert 0 \rangle =
\; e^{-\frac{1}{2} {\lVert f+g \rVert}^2}. \label{vAhh}
\end{equation}

Now, one can consider the algebra ${\cal W}({\cal M})$ obtained by taking products and linear combination of all Weyl operators defined on ${\cal M}({\cal O})$~\cite{Summers:1987fn,Summ}.
It is known that ${\cal W}({\cal M})$ is a Von Neumann algebra~\footnote{More precisely, ${\cal W}({\cal M})$ is a *-algebra of bounded operators on a Hilbert space containing the identity operator and closed under weak topology.}. The main definitions and properties can be found in App.~\eqref{A}; the interested reader can find more details in Ref.~\cite{BR}.

Another tool that will be relevant for this work is the Reeh-Schlieder theorem~\cite{Witten:2018zxz,Haag:1992hx}. It  states that the vacuum state $ \vert 0 \rangle$ is {\it cyclic} and {\it separating} for the algebra ${\cal W}({\cal M})$, that is:  i) the set of states $\{A \vert 0 \rangle$, $ A \in {\cal W}({\cal M})\}$ is {\it dense} in the Hilbert space; ii) the condition $\{A \vert 0 \rangle = 0, A \in {\cal W}({\cal M})\}$ implies $A = 0$.

Finally, we introduce the notion of {\it commutant} 
${\cal W'}({\cal M})$ for a given Von Neumann algebra ${\cal W}({\cal M})$ as 
\begin{equation} 
	{\cal W'}({\cal M}) = \{ w';  \;\;w'\;w = w \;w'\; \;\; \forall \; w \in {\cal W}({\cal M}) \}, \label{comm} 
\end{equation} 
{\it i.e.}, ${\cal W'}({\cal M})$ contains all elements which commute with each element of ${\cal W}({\cal M})$. The so-called Haag's duality can thus be stated as~\cite{Witten:2018zxz,Haag:1992hx}:
\begin{equation}
	{\cal W'}({\cal M}) = {\cal W}({\cal M'}), \label{Hd}
\end{equation}
namely, the commutant ${\cal W'}({\cal M})$ coincides with the elements of ${\cal W}({\cal M'})$ obtained by taking test functions belonging to the symplectic complement ${\cal M'}$ of ${\cal M}$. To our knowledge, for a generic interacting Quantum Field Theory, Haag's duality is still a conjecture. Though, for free Bose fields, it has been proven in \cite{eck}.

\section{The Bell-CHSH inequality}\label{Sect4}	
	
We are now ready to face the issue of constructing the Bell-CHSH operator in Quantum Field Theory. In the sequence, the operators $(A_1,A_2)$ will be referred to as Alice's operators, while $(B_1,B_2)$ as Bob's operators. We follow here the setup outlined in~\cite{Summers:1987fn,Summ,Summers:1987ze} and introduce the notion of {\it eligibility}. The set of four field operators $(A_1, A_2, B_1, B_2)$ is called {\it eligible} for the Bell-CHSH inequality if: i) they are Hermitian, that is, $A_i=A_i^{\dagger},  B_i=B_i^{\dagger}$; ii)  they are bounded operators, taking values in the interval $[-1,1]$; iii) Alice's operators commute with Bob's operators, namely, $ \left[ A_i, B_j \right] = 0$.
The physical meaning of this last conditions is that the operators $(A_1, A_2)$ and  $(B_1, B_2)$ are localized in spacelike separated regions, so measurements performed by Alice cannot have any consequence on those performed by Bob. 

Once Alice and Bob operators have been introduced, one can define the Bell-CHSH operator 
\begin{equation} \label{BCHSH-op} 
{\cal C} = (A_1 + A_2) B_1 + (A_1 -A_2) B_2, 
\end{equation} 
and evaluate its expectation value in the vacuum state. One speaks of a violation of the Bell-CHSH inequality whenever we have~\cite{Summers:1987fn,Summ}:
\begin{equation} 
\big\vert \langle 0 \vert {\cal C} \vert 0 \rangle \big\vert > 2. \label{BCHSH-viol}
\end{equation} 

Let us introduce Alice's operators ${\cal A}_{f}=e^{i\varphi(f)}$ and ${\cal A}_{f'}= e^{i\varphi(f')}$ associated with the test functions $f, f' \in {\cal M}$ and Bob's operators ${\cal A}_{g}=e^{i\varphi(g)}$ and ${\cal A}_{g'}=e^{i\varphi(g')}$ associated with test functions $g, g' \in {\cal M}'$. Thus, using these Weyl operators, one might write for the Bell-CHSH correlator\footnote{Use has been made of ${\cal A}_f {\cal A}_g = {\cal A}_{(f+g)} $, which holds for test functions $(f,g)$ whose supports are spacelike.}
\begin{equation} 
	\langle 0 \vert {\cal C}  \vert 0 \rangle = \langle 0 \vert   {\cal A}_{(f+g)} + {\cal A}_{(f'+g)} + {\cal A}_{(f+g')} - {\cal A}_{(f'+g')} \vert 0 \rangle. \label{BLC}
\end{equation} 
One could object that, even if $\langle 0 \vert {\cal C}  \vert 0\rangle $ is real, the operator ${\cal C}$ is not manifestly Hermitian. This can be easily managed by introducing the Hermitian combination
\begin{equation} 
	{\cal {\hat A}}_{(f+g)} = \frac{\left( {\cal { A}}_{(f+g)}+ {\cal { A}}_{(f+g)}^{\dagger}\right)}{2} \;. \label{HC}
\end{equation} 
It is worth observing that, as required by the eligibility criterion, the operator ${\cal {\hat A}}_{(f+g)}$ is bounded\footnote{We remind that the norm of an operator ${\cal Q}$ acting on a Hilbert space $\cal H$ is defined as~\cite{BR} 
	\begin{equation} 
		\lVert {\cal Q} \rVert = sup \left\{  \frac{\lVert {\cal Q}x \rVert}{ \lVert x \rVert} \;; \forall x \neq 0\;, x \in {\cal H} \right\}. \label{normoperator}
\end{equation}} 
by 1, {\it i.e.},
\begin{equation}
	\lVert {\cal {\hat A}}_{(f+g)} \rVert \le \frac{1}{2} \left( \lVert{\cal { A}}_{(f+g)}\rVert + \lVert {\cal { A}}_{(f+g)}^{\dagger} \rVert \right) = 1,\label{norm} 
\end{equation}
due to the fact that, being unitary, the Weyl operators have norm 1. Thus, for the Hermitian form of the Bell-CHSH correlator,  we get 
\begin{align}
	\langle 0\vert {\cal C}  \vert 0\rangle = \langle 0 \vert  {\cal {\hat A}}_{(f+g)} + {\cal {\hat A}}_{(f'+g)} + {\cal {\hat A}}_{(f+g')} - {\cal {\hat A}}_{(f'+g')}  \vert 0 \rangle.
  \label{BLCF}
\end{align} 
After a quick computation it follows that 
\begin{align} 
	\langle 0\vert {\cal C}  \vert 0\rangle \!=\! e^{-\frac{1}{2} \lVert f+g \rVert^2} \!\!\!+\! e^{-\frac{1}{2} \lVert f'+g \rVert^2} \!\!\!+\! e^{-\frac{1}{2} \lVert f+g' \rVert^2} \!\!\!-\! e^{-\frac{1}{2} \lVert f'+g' \rVert^2}. \label{fjf}
\end{align}

In the following, we shall give a further characterization of expression  \eqref{fjf}  by  emploing the Reeh-Schlieder theorem~\cite{Witten:2018zxz,Haag:1992hx} as well as Haag's duality~\cite{Haag:1992hx,eck} and the Tomita-Takesaki theory~\cite{BR}.

\section{Tomita-Takesaki construction}\label{Sect5}

Let us proceed with the Tomita-Takesaki construction, amounting to the introduction of an anti-linear unbounded operator $S$ \cite{BR} acting on the Von Neumann algebra ${\cal W}({\cal M})$ as 
\begin{align} 
	S \; w \vert 0 \rangle = w^{\dagger} \vert 0 \rangle, \qquad \forall w \in {\cal W}({\cal M})\;,  \label{TT1}
\end{align}  
from which it follows that $S \vert 0 \rangle = \vert 0 \rangle$ and $S^2 = 1$. The operator $S$ has a unique polar decomposition~\cite{BR}
\begin{equation} 
S = J \; \Delta^{1/2} \;, \label{PD}
\end{equation} 
where $J$ is anti-linear and $\Delta$ self-adjoint and positive. The following set of properties holds~\cite{BR} 
\begin{align} 
	\Delta &= S^{\dagger} S, \quad	J \Delta^{1/2} J = \Delta^{-1/2}, \nonumber \\
	 J^2 &= 1, \quad \quad \quad \quad \,\,\,	S^{\dagger} = J \Delta^{-1/2}, \nonumber \\
	J^{\dagger} &= J, \quad \quad \quad \,\,\,\, \Delta^{-1} = S S^{\dagger}. \label{TTP}
\end{align}
The  Tomita-Takesaki theorem states that~\cite{BR}: i) $J \;{\cal W}({\cal M})\; J  =  {\cal W'}({\cal M})$ as well as $J \;{\cal W'}({\cal M}) \;J = {\cal W}({\cal M})$;\\ ii) there is  a one-parameter family of operators $\Delta^{it}$ with $t \in \mathbb{R}$ which leave ${\cal W}({\cal M})$ invariant, that is, such that the following equation holds: $\Delta^{it} \; \;{\cal W}({\cal M}) \; \Delta^{-it} = \;{\cal W}({\cal M})$.

This theorem has far reaching consequences and finds applications in many areas~\cite{Summers:2003tf}. As far as the Bell-CHSH inequality is concerned, it provides a powerful way of obtaining Bob's operators from Alice's ones by means of the anti-unitary operator $J$. The construction goes as follows. Pick up two test function $(f,f')\in {\cal M}({\cal O})$ and consider two Alice's operators $({\cal A}_f, {\cal A}_{f'})$:
\begin{equation} 
 {\cal A}_f = e^{i {\varphi}(f) } \;, \qquad  {\cal A}_{f'} = e^{i {\varphi}(f') } \;. \label{AAop}
\end{equation} 
Thus, for Bob's operators we write 
\begin{equation} 
J {\cal A}_f J \;, \qquad J {\cal A}_{f'} J \;. \label{Bobbop}
\end{equation} 
From the Tomita-Takesaki theorem, it follows that the two set of operators $({\cal A}_f, A_{f'})$ and $(J{\cal A}_f J, J {\cal A}_{f'}J)$ fulfill the necessary requirements since $(J{\cal A}_f J, J {\cal A}_{f'}J)$ belong to the commutant ${\cal W'}({\cal M})$. 

At this stage, it is worth underlining that, as shown\footnote{See, for instance,  Lemma 6 of \cite{eck}.} in \cite{eck,RD}, the action of the operators $J$ and $\Delta$ may be lifted directly into the space of the test functions, giving rise to an analogue of the Tomita-Takesaki construction:  
\begin{equation} 
J {\cal A}_f J =  J e^{i {\varphi}(f) } J  \equiv e^{-i {\varphi}(jf) } \;, \label{jop}
\end{equation} 
where the test functions $jf \in {\cal M'}$. Analogously, 
\begin{equation} 
\Delta^{1/2} {\cal A}_f  \Delta^{-1/2} \equiv e^{i {\varphi}(\delta^{1/2}f) } \;. \label{dp}
\end{equation} 
The operators $(j,\delta)$ are such that \cite{eck,RD,Summ}
\begin{eqnarray}
s & = & j \delta^{1/2}, \qquad \,\, s^2=1, \nonumber \\
j \delta^{1/2} j & = & \delta^{-1/2}, \nonumber  \\
s^{\dagger} & = & j \delta^{-1/2}, \quad s^{\dagger} s^{\dagger} = 1. \label{ssd}
\end{eqnarray}
The operator $j$ is anti-unitary while $\delta$ is self-adjoint and positive. 
Moreover, in the case in which the region ${\cal O}$ is a wedge region of the Minkowski spacetime, the spectrum of the operator $\delta$ consists of a continuous spectrum $\sigma_c$ (see Appendix~\ref{B}), given by the positive real line, {\it i.e.}, $\log(\delta) =\mathbb{R}$~\cite{Summ}. This result follows from the Bisognano-Wichmann~\cite{BW} analysis of the Tomita-Takesaki modular operator for wedge regions. In what follows we shall always consider the region ${\cal O} $ to be a wedge region. For instance, one may figure out that Alice is located in the left Rindler wedge, while Bob is in the right one.

The operators $(s,s^{\dagger})$ have the physical meaning of projecting into the space ${\cal M}$ and its symplectic complement ${\cal M'}$, namely, one can show \cite{RD, Summ}  that\footnote{See Sect.2 and Proposition 4.1 of \cite{RD}. }  a test function $f$ belongs to $\cal M$ if and only if
\begin{equation} 
s f = f \;. \label{sf}
\end{equation} 
Analogously, a test function $g$ belongs to the symplectic complement ${\cal M'}$ if and only if
\begin{equation} 
s^{\dagger} g = g \;. \label{sdg}
\end{equation} 
Let us provide a simple understanding of Eqs.\eqref{sf},\eqref{sdg}. We start from equations  \eqref{TT1}. Thus, 
\begin{equation} 
S e^{i {\varphi}(f) } \vert0\rangle = e^{-i {\varphi}(f) } \vert0\rangle = S e^{i {\varphi}(f) } S \vert0\rangle \;. \label{sl1}
\end{equation} 
Furthermore,
\begin{eqnarray} 
S e^{i {\varphi}(f) } S \vert0\rangle & = & J \Delta^{1/2} e^{i {\varphi}(f) } J \Delta^{1/2} \vert0\rangle = J \Delta^{1/2}  e^{i {\varphi}(f) } \Delta^{-1/2} J \vert0\rangle \nonumber \\
& = & J e^{i {\varphi}(\delta^{1/2}f) } J  \vert0\rangle =  e^{-i {\varphi}(j \delta^{1/2} f)} \vert0\rangle  \nonumber \\
& = & e^{-i {\varphi}(f)} \vert0\rangle
  \;, \label{sl2}
\end{eqnarray} 
providing thus a comprehension of Eq.\eqref{sf}. Let us also prove that, if $sf=f$, then $s^{\dagger} (jf) = (jf)$. In fact, one has 
\begin{equation}
s^{\dagger} (jf) =  \delta^{1/2} j j f = \delta^{1/2} f = jj \delta^{1/2} f = j sf = (jf)\;. \label{sl5}
\end{equation}

Therefore, we achieved  a way of obtaining Bob's operators from Alice's ones by means of the anti-unitary operator $J$. It remains to be shown that it is always possible to adopt a set of test functions $(f,f',jf,jf')$ leading to a violation of the Bell-CHSH inequality. This will be the task of the next section. 

\section{ Bell-CHSH inequality violations }\label{Sect6}	
	
In order to obtain a Bell-CHSH inequality violation, it is necessary to perform a careful choice of the test functions $(f,f',jf, jf')$, and the Tomita-Takesaki setup encoded in the operators $(j, \delta)$ is the fundamental tool to achieve this~\cite{Summ}.

Since the operator $\delta$ has only continuous spectrum, we cannot speak of eigenvalues and eigenstates. In such a case one  introduces the notion of spectral subspaces and of approximate eigenvalues and eigenvectors, see Appendix~\ref{B}. Following~\cite{Summ}, we introduce then the spectral subspace of the operator $\delta$ specified by the interval $\left[ \lambda^2 -\varepsilon, \lambda^2+ \varepsilon \right] \subset (0,1)$. Let us denote by $\phi$ a normalized vector belonging to this subspace. As observed in~\cite{Summ}, it turns out that
\begin{equation} 
\delta (j \phi) = j (\delta^{-1} \phi) \;, \qquad \delta^{-1} (j \phi) = j (\delta \phi) \; \label{exch}
\end{equation} 
showing that the modular conjugation $j$ exchanges the spectral subspace $\left[  \lambda^2 -\varepsilon, \lambda^2+ \varepsilon \right]$ into $\left[  \frac{1}{\lambda^2} -\varepsilon, \frac{1}{\lambda^2}+ \varepsilon \right]$. As a consequence, $\phi$ and $j \phi$ are orthogonal:
\begin{equation} 
\langle \phi \vert j \phi \rangle =0 \;. \label{orth}
\end{equation} 
Let us proceed now with the introduction of Alice's test functions $(f,f')\in {\cal M}$. Following~\cite{Summ}, we set
\begin{eqnarray} 
f & = & \eta (1 + s) \phi \;, \nonumber \\
f' & = & \eta'(1+s) (i\phi) \;, \label{ffA}
\end{eqnarray} 
where $(\eta, \eta')$ are two real arbitrary positive constants which, as much as the Bell-CHSH angles of Quantum Mechanics, will be chosen at our best convenience. The presence of the operator $(1+s)$ in Eq.~\eqref{ffA} ensures that both $f$ and $f'$ belong to ${\cal M}$: 
\begin{equation} 
s f=f \;, \qquad s f'= f'  \;. \label{sff}
\end{equation}
Let us evaluate the norm of $f$\footnote{In evaluating $\vert \vert f \vert \vert^2$, one has to remind that, in the case of an anti-unitary operator $\cal U$, one has $\langle {\cal U}x \vert {\cal U}y \rangle = \langle y \vert x \rangle$.}
\begin{align} 
\vert \vert f \vert \vert^2 &= \eta^2 \langle (1+s)\phi \; \vert \;(1+s)\phi \rangle  \nonumber \\
&= \eta^2 \left( 1 + \langle j \delta^{1/2} \phi \vert j \delta^{1/2} \phi \rangle \right) = \eta^2 (1 +\mu) \;, \label{normf}
\end{align} 
where we have introduced
\begin{equation} 
\mu = \langle \phi | \;\delta \phi \rangle \;. \label{muq}
\end{equation}
Since $\phi$ belongs to the spectral subspace $\left[  \lambda^2 -\varepsilon, \lambda^2+ \varepsilon \right]$, the quantity $\mu$ depends on $\lambda^2$ and, in the limit $\varepsilon \rightarrow 0$, one has $\mu \rightarrow \lambda^2$. In much the same way 
\begin{equation} 
\vert \vert f' \vert \vert^{2} = {\eta'}^{2} (1 + \mu) \label{f1f1norm} \;. 
\end{equation} 
From expressions \eqref{normf} and \eqref{f1f1norm} one sees that the parameters $(\eta, \eta')$ are related to the norm of $(f,f')$. It is important to realize that the possibility of treating $(\eta,\eta')$ as free parameters is a consequence of the direct use of the Weyl operators in the whole analysis of the Bell-CHSH inequality. In fact, the unitarity property of the Weyl operators holds for any test function $ f \in {\cal M}$, {\it i.e.},
\begin{equation} 
{\cal A}_f = e^{i \varphi(f)}, \qquad {\cal A}_f^{\dagger} {\cal A}_f = {\cal A}_f {\cal A}_f^{\dagger} = 1, \;\;\; \forall f \in {\cal M}. \label{unit}
\end{equation} 
As the operators entering the Bell-CHSH inequality are Hermitian combinations of the Weyl operators, it turns out that they remain bounded by 1 for any choice of the norm of the test functions. Therefore, we can consider the parameters $(\eta, \eta')$ as free parameters.

\begin{figure}[t!]
	\begin{minipage}[b]{1.0\linewidth}
		\includegraphics[width=\textwidth]{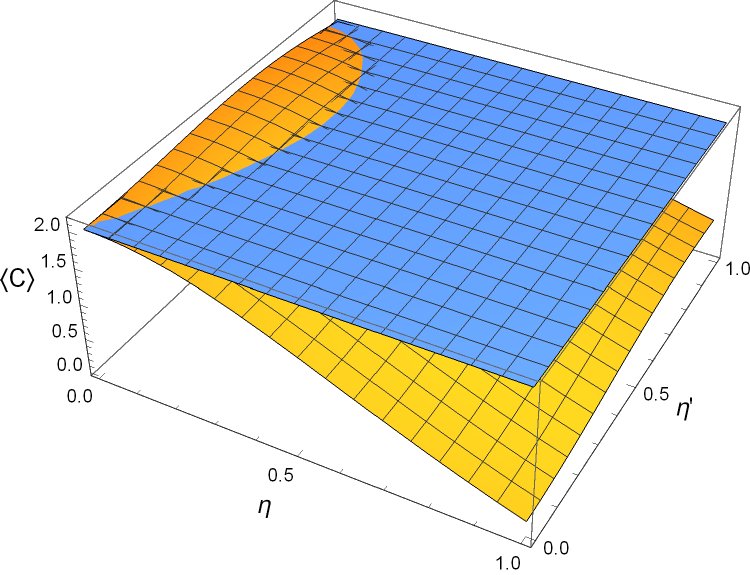}
	\end{minipage} \hfill
	\begin{minipage}[b]{1.0\linewidth}
	\includegraphics[width=\textwidth]{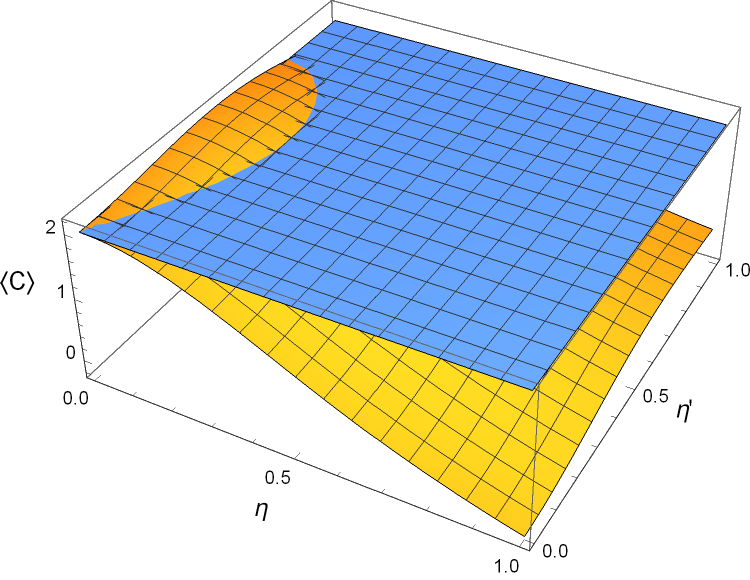}
\end{minipage} \hfill
\caption{Behavior of the Bell-CHSH correlator $\langle 0 \vert {\cal C} \vert 0 \rangle $ as a function of the parameters $(\eta, \eta') \in [0,1]$ for $\lambda=0.5$ (top panel) and $\lambda=0.99$ (lower panel). The violation corresponds to the orange region above the blue plane $\langle 0 \vert {\cal C} \vert 0 \rangle =2$.}
	\label{lambda05}
	\end{figure}

We turn now to $(j f, jf')$. For the norms $\vert \vert jf \vert \vert^2$ and $\vert \vert jf' \vert \vert^2$ we immediately get 
\begin{equation} 
\vert \vert jf \vert \vert^2 = \eta^2(1+ \mu) \;, \qquad \vert \vert jf' \vert \vert^2 = {\eta'}^2 (1 + \mu) \;. \label{jfjff}
\end{equation}
Let us evaluate $\langle f \vert jf \rangle$: 
\begin{equation} 
\langle f \vert jf \rangle = \eta^2 \left( \langle (1 + j\delta^{1/2})\phi \; \vert\; (j\phi + \delta^{1/2} \phi \rangle   \right) = 2 \eta^2 \mu^{1/2}. \label{fjfa1}
\end{equation}
Similarly,
\begin{equation} 
\langle f \vert jf' \rangle = 0 \;, \qquad \langle f' \vert jf' \rangle = 2 \mu^{1/2} {{\eta'}}^2. \label{fip}
\end{equation} 
Therefore, taking the limit $\varepsilon \rightarrow 0$, for the Bell-CHSH correlator, we get the rather simple expression 
\begin{equation} 
\langle 0\vert {\cal C} \vert0\rangle = e^{-\eta^2(1+ \lambda)^2} + 2 e^{-\frac{1}{2}(1+\lambda^2) (\eta^2 + {\eta'}^2)}  - e^{- {\eta'}^2 (1+ \lambda)^2}, \label{Cf1}
\end{equation}
where, as previously underlined, $(\eta,\eta')$ are free parameters. To infer that $\langle 0\vert {\cal C} \vert0\rangle$ exhibits a violation of the classical bound, we have to show that it is possible to find non-vanishing values of $(\eta, \eta') \neq 0$ such that
 \begin{equation} 
 \vert \langle 0\vert {\cal C} \vert0\rangle \vert > 2. \label{pv} 
 \end{equation} 
To check out that this is indeed the case, we have performed a numerical study of expression \eqref{Cf1}, see Fig.~\ref{lambda05}. 

The blue plane corresponds to the classical bound  $ \vert \langle 0 \vert {\cal C} \vert 0\rangle \vert =2$, while the orange surface shows the behavior of  $\vert \langle 0\vert {\cal C} \vert0\rangle \vert $ as a function of $(\eta, \eta')$. Despite the presence of rapidly decreasing exponentials in Eq.~\eqref{Cf1}, Fig.~\ref{lambda05}  shows a good range of values for $(\eta, \eta')$ yielding the Bell-CHSH inequality violation. More precisely, Fig.~\ref{lambda05} exhibits the Bell-CHSH correlator behavior as a function of the free parameters $(\eta, \eta') \in [0,1]$  for the cases $\lambda=0.5$ and $0.99$. One observes the existence of a region for which the orange surface is above the plane $\langle 0 \vert {\cal C} \vert 0 \rangle =2$, explicitly exhibiting the Bell-CHSH inequality violation.

\section{Conclusions}\label{Sect7}

In this work we have discussed the violation of the Bell-CHSH inequality in the vacuum state of a relativistic free scalar Quantum Field Theory. The violation has been achieved by direct evaluation of the correlation functions of Weyl operators. Needless to say, these operators display a remarkable structure, Eq.~\eqref{algebra}, encoding the causality properties of the spacetime. 

It is worth highlighting the differences with the pionering works~\cite{Summers:1987fn,Summ}, where the Bell-CHSH inequality violation for the scalar field is achieved by mapping the whole problem into a Quantum Mechanical system of a pair of oscillators whose vacuum state is an entangled squeezed state. The Bell operators are thus obtained by making use of the pseudospin operators~\cite{Larsson:2003pjo} acting on the Hilbert space of the two oscillators system. Instead, here we have avoided the use of a mapping with Quantum Mechanics, relying entirely on the algebraic properties of Weyl operators. Although our results for the Bell-CHSH inequality violation are not close to Tsirelson's bound~\cite{tsi1}, {\it i.e.}, $2 \sqrt{2}$, we point out the high simplicity of the Bell-CHSH correlator computation, as shown by the final expression, Eq.~\eqref{Cf1}, which enables one to detect a region in the parameter space for which the Bell-CHSH inequality violation is clearly visible, see Fig.~\ref{lambda05}.

Let us also underline that the use of Weyl operators, combined with Tomita-Takesaki construction, opens the route to generalize the present results to the quantum field theory version of the Mermin inequalities~\cite{Mermin}  for scalar fields. In fact, the spectrum properties~\cite{Summ} of the modular operator $\delta$, Eq.~\eqref{ssd}, allow us to introduce a sequence of spectral subspaces ${\cal I}_n$, $n=1,2,3,...$, and corresponding normalized vectors  $\phi_{n}$. Moreover, acting with the projector $(1+s)/2$ on $\phi_{n}$ we can construct a sequence of test functions $(f_{n}, f'_{n})$ localized in the right Rindler wedge. Therefore, from the Tomita-Takesaki setup, it follows that the test functions $(jf_{n}, jf'_{n})$ are localized in the left Rindler wedge. Employing then the powerful property of the correlation function of $n$ spacelike separated Weyl operators, as expressed by
\begin{equation} 
\langle 0 \vert  {\cal A}_{h_1} \;.....\; {\cal A}_{h_n} \vert 0 \rangle = e^{-\frac{1}{2} \lVert h_1 +... + h_n \rVert^2}, \label{nWeyl} 
\end{equation} 
the Mermin inequality of order $n$ can be faced. We shall report on this nice application of Weyl operators and the corresponding Tomita-Takesaki framework in a forthcoming work. 
	
\section*{Acknowledgments}
	The authors would like to thank the Brazilian agencies CNPq,  FAPERJ  and CAPES, for financial support.  S.P.~Sorella, I.~Roditi, and M.S.~Guimaraes are CNPq researchers under contracts 301030/2019-7, 311876/2021-8, and 310049/2020-2, respectively. G. Peruzzo is a FAPERJ postdoctoral fellow in the {\it Pós-Doutorado Nota 10} program under contracts E-26/205.924/2022 and E-26/205.925/2022.


\appendix

\section{Basic notions on Von Neumann algebras and Tomita-Talesaki theory}\label{A}

For the benefit of the reader, we provide here a rather concise summary on some basic notions on Von Neumann algebras and on the Tomita-Takesaki theory. The material is by no means to be considered exhaustive. The reader can consult the classic book~\cite{BR} for more details and proofs of the statements.

\subsection{Algebras, $^*$-algebras, $C^*$-algebras}

Let ${\cal F}$ be a vector space over $\mathbb{C}$. ${\cal F}$ is called an algebra if it is equipped with a multiplication law which associates the product $AB$ to each pair $(A,B) \in {\cal F}$. The product is assumed to be associative and distributive
\begin{eqnarray} 
A (BC) & =& (AB)C \;, \nonumber \\
A(B+C) & = & AB + AC \; \nonumber \\
\alpha \beta (AB) & =& (\alpha A ) (\beta B) \;, \qquad (A,B) \in {\cal F}\;, (\alpha, \beta) \in \mathbb{C} \;. \label{ass}
\end{eqnarray} 
The algebra ${\cal F}$ is called commutative or Abelian, if the product is commutative: $(AB) = (BA)$.\\\\\underline{\bf Involution} A mapping $A\in {\cal F}$ $\rightarrow$ $A^*\in {\cal F}$ is called an involution, or adjoint operation, if it has the following properties: 
\begin{eqnarray} 
(A^*)^* & = & A \;, \nonumber \\
(AB)^* & = & B^* A^* \;, \nonumber \\
(\alpha A^* + \beta B^*) & = & \overline{\alpha}  A^* + {\overline \beta} B^*  \;, \label{inv} 
\end{eqnarray} 
where $\overline{\alpha}, \overline{\beta}$ are the complex conjugates of $\alpha, \beta$. \\\\\underline{\bf Normed Algebras}  The algebra $\cal F$ is a normed algebra if a norm $||\;\;||: {\cal F} \rightarrow \mathbb{R}$ is assigned: 
\begin{eqnarray} 
|| A|| &  \ge & 0 \;, \qquad ||A||= 0 \Rightarrow A=0 \;, \nonumber \\
|| A + B|| & \le & ||A|| + ||B|| \;, \nonumber \\
|| A B|| & \le & ||A|| ||B||  \;. \label{normalg}
\end{eqnarray} 
\underline{\bf Banach Algebra} If the normed algebra ${\cal F}$ is complete with respect to the norm $||\;\;||$, {\it i.e.}, if every Cauchy sequence converges to an element of $\cal F$, ${\cal F} $ is called a Banach algebra. \\\\\underline{\bf Banach *-algebra}  A normed complete algebra equipped with an involution $*$ and such that 
\begin{equation} 
|| A^* || = ||A|| \;, \label{stB}
\end{equation}
is called a Banach * -algebra. \\\\\underline{\bf $C^*$-algebra} A $C^*$-algebra is a Banach *-algebra with the property 
\begin{equation} 
|| A^* A || = ||A||^2  \qquad \forall A \in {\cal F} \;. \label{starcond}
\end{equation}
This property is called the *-condition. The *-condition automatically implies that $||A^*|| = ||A||$.

\subsection{Convergence in ${\cal B}({\cal H})$} 
Let ${\cal H}$ be a Hilbert space and let ${\cal B}({\cal H})$ denote the set of all bounded operators acting on 
${\cal H}$. Several topologies can be defined on ${\cal B}({\cal H})$, namely 
\begin{itemize} 
\item \underline{ Norm topology.} It is defined by the norm operator 
\begin{equation} 
|| {\cal A} || = sup \left\{  \frac{||{\cal A}\psi||}{||\psi||} \;; \forall \psi \neq 0\;, \psi \in {\cal H} \right\} \;. \label{normA}
\end{equation}
A sequence of operators $(A_i)_{i \in \mathbb{N}} \in {\cal B}({\cal H})$  converges in the norm topology to $A\in {\cal B}({\cal H})$ if $||A_i -A||$ converges to zero 
\begin{equation} 
\textrm{ lim}_{i \rightarrow \infty}  ||A_i -A|| = 0 \;. \label{normconv} 
\end{equation} 
\item \underline{Strong topology.} It is defined by the semi-norm $\nu_\psi(A)$ 
\begin{equation}
\nu_\psi(A)  = ||A \psi || \;, \qquad \psi\in {\cal H} \label{strt} 
\end{equation}
A sequence of operators $(A_i)_{i \in \mathbb{N}} \in {\cal B}({\cal H})$  converges in the strong  topology to $A\in {\cal B}({\cal H})$ if $||A_i \psi -A\psi ||$ converges to zero 
\begin{equation} 
\textrm{ lim}_{i \rightarrow \infty}  ||A_i \psi -A\psi || = 0 \;, \forall \psi \in {\cal H}  \;. \label{strnormconv} 
\end{equation} 
\item \underline{Weak  topology.} It is defined by the semi-norm $\nu_{\psi\varphi}(A)$ 
\begin{equation}
\nu_{\psi\varphi}(A)  = \langle \varphi \; | \; A \psi \rangle \;, \qquad \varphi, \psi \in {\cal H} \label{wtrt} 
\end{equation}
A sequence of operators $(A_i)_{i \in \mathbb{N}} \in {\cal B}({\cal H})$  converges in the weak  topology to $A\in {\cal B}({\cal H})$ if 
\begin{equation} 
\textrm{ lim}_{i \rightarrow \infty}  | \langle \varphi \; | \; A_i \psi -A\psi \rangle | = 0 \;, \forall \varphi, \psi \in {\cal H}  \;. \label{wtrnormconv} 
\end{equation} 
\item From the Cauchy-Schwartz inequality and from the properties of the norm operator, it follows that the convergence given by the norm topology is the strongest one. Norm convergence implies strong convergence which, in turn, implies weak convergence. \\\\The pair $\left( {\cal B}({\cal H}), ||\;\;||\right) $ where $||\;\;||$ is the norm operator, Eq.\eqref{normA}, is a $C^*$-algebra. 
\end{itemize}

\subsection{Von Neumann algebras} 

\underline{\bf The commutant} For any subset ${\cal F} \subset {\cal B}({\cal H}) $, the commutant ${\cal F'}$ is the set of all operators of ${\cal B}({\cal H})$ which commute with each element of ${\cal F}$ 
\begin{equation} 
{\cal F'} = \{ B \in {\cal B}({\cal H}) \; \textrm{ such that} \; AB=BA\;, \forall A\in {\cal F} \}  \;. \label{comm}
\end{equation} 
Von Neumann algebras can be defined in two equivalent ways.\\\\\underline{\bf Definition} A *-subalgebra ${\cal F} \in {\cal B}({\cal H})$ containing the identity is a Von Neumann algebra if it is closed under weak topology. \\\\\underline{\bf Definition} A *-subalgebra ${\cal F} \in {\cal B}({\cal H})$ containing the identity is a Von-Neumann algebra if ${\cal F}$ is such that
\begin{equation} 
{\cal F} = ({\cal F'})' \;. \label{MMp}
\end{equation}
The equivalence between these two definitions is ensured by Von Neumann's bicommutant Theorem:\\\\\underline{\bf Bicommutant Theorem}  Let   ${\cal F}$ be a *-subalgebra of  ${\cal B}({\cal H}) $. The following conditions are equivalent
\begin{eqnarray} 
i) \;\;\;& {\cal F} & =  ({\cal F'})'  \;, \nonumber \\
ii) \;\;\;& {\cal F} & \;\;\textrm{is closed under weak topology} \;, \nonumber \\
iii) \;\;\;& {\cal F} &\;\; \textrm{is closed under strong topology} \;. \label{bicomm} 
\end{eqnarray} 
\underline{\bf Definition} A vector $ \Omega \in {\cal H}$ is a cyclic vector for the Von Neumann algebra ${\cal F}$ if 
\begin{equation} 
 \textrm {span} \{ {\cal F} \Omega\} \; \textrm{is dense in} \;{\cal H}. \;. \label{cycl}
\end{equation} 
\underline{\bf Definition} A vector $ \Omega \in {\cal H}$ is called separating for a Von Neumann algebra ${\cal F}$ if the condition $A \Omega=0$, $A \in {\cal F}$ implies that $A=0$. \\\\\underline{\bf Remark} Let ${\cal F}$ be a Von Neumann algebra and $\Omega \in {\cal H}$. The following statements are equivalent:
\begin{eqnarray} 
i) \;  \Omega &\; & \textrm{is cyclic for} \; {\cal F} \;, \nonumber \\
ii) \; \Omega &\; & \textrm{is separating  for} \; {\cal F'} \;.  \label{mmm}
\end{eqnarray} 
\subsection{Modular structure and the Tomita-Takesaki Theorem} 
 
\underline{\bf Definition of the anti-linear operators $S$ and $S^\dagger$} 

\noindent Let $\Omega \in {\cal H}$ be a separating and cyclic vector for the Von Neumann algebra $\cal F$, then $\Omega$ is also separating and cyclic for the commutant ${\cal F'}$. Define the operators 
\begin{eqnarray} 
& \bullet &\;\; S\; A\; \Omega = A^* \; \Omega  \;, \nonumber \\
& \bullet &\;\; S^{\dagger}\; A'\; \Omega = (A')^* \; \Omega \;. \label{SSd}
\end{eqnarray} 
The operators $(S,S^{\dagger})$ are anti-linear and 
\begin{equation} 
SS = S^{\dagger} S^{\dagger} = 1 \qquad S \Omega = S^{\dagger} \Omega = \Omega\;. \label{SSD1}
\end{equation} 
\\
\underline{\bf Polar decomposition} 
\begin{equation} 
S = J \Delta^{1/2} \;, \label{PD}
\end{equation} 
where $J$ is an anti-linear operator, called the conjugation operator. $\Delta$ is a self-adjoint, positive operator called the modular operator. The following properties hold
\begin{eqnarray} 
J \Delta^{1/2} J & = & \Delta^{-1/2}  \;, \nonumber \\
\Delta & = & S^{\dagger} S \;, \qquad
\Delta^{-1}  =  S S^{\dagger} \;, \nonumber \\
J^{\dagger} & = & J \;, \qquad J^2 = 1 \;. \label{DJprop} 
\end{eqnarray} 
\underline{\bf Tomita-Takesaki Theorem} 

\noindent Let ${\cal F}$ be a Von Neumann algebra with a cyclic and separating vector $\Omega \in {\cal H}$. Let $\Delta$ be the associated modular operator and $J$ the conjugation operator. Then 
\begin{eqnarray} 
& \bullet & \;\;\; J {\cal F}J = {\cal F'} \;, \qquad J {\cal F'} J = {\cal F}  \;, \nonumber \\
& \bullet & \;\;\; \textrm{there exists a one parameter family of unitary operators} \nonumber \\ & & \Delta^{it}, \; t \in \mathbb{R},  \textrm{such that}\;\;   \Delta^{it} {\cal F} \Delta^{-it} = {\cal F} \;. \label{TTtheorem}
\end{eqnarray} 

\section{Short account on the spectral properties of self-adjoint operators}\label{B}

Let us give here a short account on a few notions of the spectral theory. For more details, see Refs.~\cite{BR,spectral}.

\subsection{ Resolvent and spectrum} 
Let $X$ be a Banach space equipped with the norm $\vert \vert \; . \; \vert \vert$ and let $T: {\cal D}(X) \rightarrow X$  a linear operator. The resolvent $\rho(T)$ is the set of all $\nu \in \mathbb{C}$ such that the operator 
\begin{equation} 
{\cal R}_\nu(T) = ( T -\nu I)^{-1} \;, \label{resolv} 
\end{equation}
fulfills the following properties:
\begin{itemize} 
\item {\it i)} ${\cal R}_\nu(T)$ exists, {\it i.e.}, $ {\rm kernel \; of}  \;{\cal R}_\nu(T) = \{0\}$;
\item {\it ii)} ${\cal R}_\nu(T)$ is bounded;
\item {\it iii)} ${\cal R}_\nu(T)$ is densely defined in $X$. 
\end{itemize}
The spectrum $\sigma(T)$ of $T$ is the complement of $\rho(T)$, namely 
\begin{equation} 
\sigma(T) = \mathbb{C}/\rho(T) \;. \label{spec}
\end{equation}
It consists of several parts: 
\begin{itemize} 
\item \underline{\bf Point spectrum $\sigma_p(T)$}: property $i)$ is not fulfilled, {\it i.e.}  $ {\rm kernel \; of} \;{\cal R}_\nu(T) \neq \{0\}$. There exist non-vanishing $x \in X$, $x \neq 0$, such that 
\begin{equation} 
T x = \nu x \;. \label{eige}
\end{equation} 
In such a case, $\nu$ is called eigenvalue and $x$ eigenvector. The point spectrum $\sigma_p(T)$ is the set of all eigenvalues. 
\item \underline{\bf Continuous spectrum $\sigma_c(T)$}. Here, properties $i)$ and $iii)$ hold. Though, $ii)$ is not fulfilled. The operator ${\cal R}_\nu(T) $ exists but is unbounded. This means that there is a sequence $\{ x_n \}_{n \in \mathbb{N}}$, $||x_n||=1$,  such that 
\begin{equation} 
||{\cal R}_\nu(T) x_n|| \rightarrow \infty \;\;\;\; {\rm for }\;\;  n \rightarrow \infty 
\end{equation} 
In this case, one introduces the sequence 
\begin{equation} 
y_n = \frac{{\cal R}_\nu(T) x_n}{||{\cal R}_\nu(T) x_n||}  \;, \label{seqy}
\end{equation}
so that 
\begin{equation} 
(T -\nu I) y_n = \frac{ x_n}{||{\cal R}_\nu(T) x_n||}  \;, \label{seqy1}
\end{equation}
from which it follows that 
\begin{equation} 
||(T -\nu I) y_n|| \rightarrow 0 \;\;\;\; {\rm for }\;\;  n \rightarrow \infty \;. \label{seqy2} 
\end{equation} 
The unit vectors $\{ y_n \}$ are referred to as approximate eigenstates. 
\item \underline{\bf Residual spectrum $\sigma_r(T)$ }. Properties $i)$ and $ii)$ hold, though $iii)$ fails. 
\end{itemize}
The spectrum $\sigma(T)$ of $T$ is thus 
\begin{equation} 
\sigma(T) = \sigma_p(T) \cup \sigma_c(T) \cup \sigma_r(T)  \;. \label{specdef}
\end{equation} 
Let us mention that the residual spectrum $\sigma_r(T)$ is absent for self-adjoint operators.

\subsection{The spectral theorem for (unbounded) self-adjoint operators} 

Finally, let us state the so-called spectral theorem:\\\\\underline{\bf Spectral Theorem for bounded  operators} 

Let $A \in {\cal B}({\cal H})$ be a self-adjoint operator. It exists a unique spectral measure on the Borel $\sigma$-algebra ${\cal B}_{Borel}(\sigma(A))$ such that 
\begin{equation} 
A = \int_{\sigma(A)} \nu dE_\nu  \;, \label{sTb}
\end{equation}
where
\begin{itemize}
\item $E_B$ is an orthogonal projector
\begin{eqnarray} 
E^2_B & = & E_B \;, \qquad E^{*}_B = E_B   \;\;\;\; \forall B \in {\cal B}_{Borel}(\sigma(A)) \;, \nonumber \\[3mm]
E_{\emptyset} & = & 0 \;, \qquad E_{\sigma(A)} = 1 \;, \nonumber \\[3mm]
E_{B_1} E_{B_2} & = & E_{B_1} \cap E_{B_2}  \;\;\;\;\forall B_1, B_2 \in {\cal B}_{Borel}(\sigma(A)) \;, \label{sp2}
\end{eqnarray} 
\item if $(B_n)_{n \in \mathbb{N}}$ is a countable collection of ${\cal B}_{Borel}(\sigma(A))$ with $B_k \cap B_l = \emptyset$, then
\begin{equation} 
E_{\cup_k B_k} x = {\rm lim}_{N\rightarrow \infty} \sum_{k=1}^{N} E_{B_n} x \;\;\;\; \forall x \in {\cal H} \;. \label{spt2} 
\end{equation} 
\end{itemize}
More generally, if $f(\nu)$ is a continuous function on $\sigma(A)$,  one has the spectral calculus
\begin{equation} 
f(A) = \int_{\sigma(A)} f(\nu) \; \;dE_\nu  \;, \label{sTb}
\end{equation} 
with 
\begin{eqnarray} 
f(A) x &  = & \int_{\sigma(A)} f(\nu) \;d(E_\nu x)  \;, \nonumber \\
||f(A) x||^2 & = & \int_{\sigma(A)} |f(\nu)|^2 \;d(\langle x\;|E_\nu|\; x\rangle )  \;, \nonumber \\
\langle x \; | f(A) |\; y \rangle & = & \int_{\sigma(A)} f(\nu) \;d(\langle x\;|E_\nu|\; y\rangle )  \;\;\;\; x,y \in {\cal H} \;.  \nonumber \\ \label{fsp}
\end{eqnarray} 
\\\\\underline{\bf Spectral Theorem for unbounded  operators} 

Let $A$ be a self-adjoint operator on ${\cal H}$. It exists a unique spectral measure on the Borel $\sigma$-algebra ${\cal B}_{Borel}(\mathbb R)$ such that 
\begin{equation} 
A = \int_{\mathbb R} \nu \;dE_\nu  \;. \label{sTub}
\end{equation}

\newpage
	

\begin{thebibliography}{99}


\bibitem{Witten:2018zxz}
E.~Witten, {\it APS Medal for Exceptional Achievement in Research: Invited article on entanglement properties of quantum field theory}, Rev. Mod. Phys. \textbf{90},  045003 (2018).

\bibitem{Faulkner:2022mlp}
T.~Faulkner, T.~Hartman, M.~Headrick, M.~Rangamani, and B.~Swingle, {\it Snowmass white paper: Quantum information in quantum field theory and quantum gravit}, arXiv:2203.07117.



\bibitem{Bell:1964kc}
J.~S.~Bell, {\it On the Einstein-Podolsky-Rosen paradox}, Physics Physique Fizika \textbf{1}, 195 (1964).



\bibitem{Clauser:1969ny}
J.~F.~Clauser, M.~A.~Horne, A.~Shimony, and R.~A.~Holt,
{\it Proposed experiment to test local hidden variable theories}, Phys. Rev. Lett. \textbf{23}, 880 (1969).

\bibitem{Fabbrichesi21}
M. Fabbrichesi, R. Floreanini, and G. Panizzo, {\it Testing Bell Inequalities at the LHC with Top-Quark Pairs}, Phys. Rev. Lett. {\bf 127}, 161801 (2021).

\bibitem{Severi22}
C. Severi, C. D. E. Boschi, F. Maltoni, and M. Sioli, {\it Quantum tops at the LHC: from entanglement to Bell inequalities}, Eur. Phys. J. C {\bf 82}, 285 (2022).

\bibitem{Afik21}
Y. Afik and J. R. M. de Nova, {\it Entanglement and quantum tomography with top quarks at the LHC}, Eur. Phys. J. Plus {\bf 136}, 907 (2021).

\bibitem{Afik22}
Y. Afik and J. R. M. de Nova, {\it Quantum information with top quarks in QCD}, Quantum {\bf 6}, 820 (2022).

\bibitem{Afik23}
Y. Afik and J. R. M. de Nova, {\it Quantum discord and steering in top quarks at the LHC}, Phys. Rev. Lett. {\bf 130}, 221801 (2023).

\bibitem{Barr22}
A.~J.~Barr, {\it Testing Bell inequalities in Higgs boson decays}, Phys. Lett. B {\bf 825},  136866 (2022).		

\bibitem{Ashby-Pickering:2022umy}
R.~Ashby-Pickering, A.~J.~Barr, and A.~Wierzchucka, {\it Quantum state tomography, entanglement detection and Bell violation prospects in weak decays of massive particles},  J. High Energ. Phys. \textbf{2023}, 020 (2023).


\bibitem{Summers:1987fn}
S.~J.~Summers and R.~Werner, {\it Bell's Inequalities and Quantum Field Theory. 1. General Setting}, J. Math. Phys. \textbf{28}, 2440 (1987).

\bibitem{Summ} 
S.~J.~Summers and R.~Werner, {\it Bell’s inequalities and quantum field theory. II. Bell’s inequalities are maximally violated in the vacuum}, J. Math. Phys. {\bf 28}, 2448 (1987).


\bibitem{Summers:1987ze}
S.~J.~Summers and R.~Werner, {\it Maximal Violation of Bell's Inequalities Is Generic in Quantum Field Theory}, Commun. Math. Phys. \textbf{110}, 247 (1987).


\bibitem{Haag:1992hx}
R.~Haag, {\it Local quantum physics: Fields, particles, algebras}, Springer-Verlag, 1992



\bibitem{Summers:1995mv}
S.~J.~Summers and R.~F.~Werner, {\it On Bell's inequalities and algebraic invariants}, Lett. Math. Phys. \textbf{33}, 321 (1995).

\bibitem{Verch} 
R.~ Verch and R.~ F.~ Werner, {\it Distillability and positivity of partial transposes in general quantum field systems}, Rev. Math. Phys. {\bf 17}, 545 (2005).


\bibitem{Peruzzo:2022pwv}
G.~Peruzzo and S.~P.~Sorella, {\it Remarks on the Clauser-Horne-Shimony-Holt inequality in relativistic quantum field theory}, Phys. Rev. D \textbf{106}, 125020 (2022).

\bibitem{Peruzzo:2022tog}
G.~Peruzzo and S.~P.~Sorella, {\it Feynman path integral formulation of the Bell-Clauser-Horne-Shimony-Holt inequality in quantum field theory}, Phys. Rev. D \textbf{107}, 105001 (2023).

\bibitem{Sorella:2023pzc}
S.~P.~Sorella, {\it Remarks on the Bell-Clauser-Horne-Shimony-Holt inequality}, Phys. Rev. D \textbf{107}, 125013 (2023).

\bibitem{DeFabritiis:2023ubj}
P.~De Fabritiis, I.~Roditi and S.~P.~Sorella, {\it Mermin's inequalities in Quantum Field Theory}, Phys. Lett. B {\bf 846}, 138198 (2023).



\bibitem{Dudal:2023pbc}
D.~Dudal, P.~De Fabritiis, M.~S.~Guimaraes, G.~Peruzzo and S.~P.~Sorella, {\it BRST invariant formulation of the Bell-CHSH inequality in gauge field theories}, accepted for publication in SciPost Physics, arXiv:2304.01028.


\bibitem{Dudal:2023mij}
D.~Dudal, P.~De Fabritiis, M.~S.~Guimaraes, I.~Roditi and S.~P.~Sorella, {\it Maximal violation of the Bell-Clauser-Horne-Shimony-Holt inequality via bumpified Haar wavelets}, Phys. Rev. D {\bf 108}, L081701 (2023).

\bibitem{BR} 
O.~Bratteli and D.~W.~Robinson, {\it Operator Algebras and Quantum Statistical Mechanics 1}, Springer 1987. 

\bibitem{eck}
I.~P.~Eckmann and K.~Osterwalder, {\it An application of Tomita's theory of modular Hilbert algebras: Duality for free Bose fields}, J. Funct. Anal. {\bf 13}, 1 (1973).

\bibitem{Summers:2003tf}
S.~J.~Summers, {\it Tomita-Takesaki modular theory}, arXiv:math-ph/0511034.

\bibitem{RD}
M.~A.~Rieffel and A.~Van Daele, {\it A bounded operator approach to Tomita-Takesaki theory}, Pacific J. Math. {\bf 69}, 187 (1977). 

\bibitem{BW}
J.~J.~Bisognano and E.~H~.Wichmann, {\it On the duality condition for a Hermitian scalar field}, J. Math. Phys. {\bf 16}, 985 (1975).

\bibitem{Larsson:2003pjo}
J.~\r{A}.~Larsson, {\it Bell Inequalities for Position Measurements}, Phys. Rev. A \textbf{70}, 022102 (2004).

\bibitem{tsi1} 
B~.S~. Cirel'son, {\it Quantum generalizations of Bell's inequality}, Lett. Math. Phys. {\bf 4}, 93 (1980). 


\bibitem{Mermin} 
N.~D.~Mermin, {\it Extreme quantum entanglement in a superposition of macroscopically distinct states},  Phys. Rev. Lett. {\bf 65}, 1838 (1990).

\bibitem{spectral}
K. Schmüdgen, {\it Unbounded self-adjoint operators on Hilbert space}, Springer Science+Business Media Dordrecht (2012).

	
\end{thebibliography}
\end{document}